# 2018 HSSA Symposium/ Friday, November, 16, 2018

## The limits of progress in the digital era

*Joaquin Luque*
University of Seville, Spain

## Abstract

The concept of *progress* clearly percolates the activities in science, technology, economy and society. It is a driving vector (probably the main vector) of our daily activity as researchers. The *#InThisGen* initiative, proudly displayed in places across the University of Berkeley's campus, and its headline lemma —what can we change in a single generation?— are clear exponents of the underlying assumption that progress is not only possible but also desirable.

But about the concept of *progress* two major concerns arise. First of all, *progress* means some kind of going forward, that is a direction in a journey. But deciding the way in the route clearly implies that we are explicit or implicitly defining the goals, as individuals and as society. That is, the concept of *progress* has a set of underlying values.

Additionally, the conceptual paradigm in scientific research —and probably in the whole spirit of our times— it is assuming some kind of endless *progress*. It is true that many technological innovations and their subsequent impact on society have found resistance, from Luddites to ecologist movements. But the last 150 years (the age of our university) have been witness of an enormous and general increase in knowledge, wealth and welfare, showing how progress can be sustained in the long-term and positively influence the human beings and the society.

The solution of the concerns about the impact of new scientific and technological advances are commonly referred to the own science and technology. The non-always-formulated assumption is that a better and more advanced science will solve the drawbacks of the nowadays science. That is, the concept of *progress* is also based on an infinite faith on the unlimited possibilities of Technology (now in uppercase).

In this contribution will try to discuss these bounds, addressing the limits of materials, scientific knowledge and technological know-how. We will mainly focus on the limitations in technological knowledge in the software design, a key aspect of the digital era. Our main thesis, which will be addressed through the paper, is that there are intrinsic limits to technological knowledge and the concept of *progress* should take them into account.

## A. The concept of progress

Perhaps one of the topics associated with the concepts of science and technology is, without doubt, that of progress. Even socially and politically the idea of progress is assumed as a positive value. But what does it mean to progress? The term "progress" derives from the Latin word *progredi* which means to walk forward. Progress is the opposite of the return. The idea of progress implies a path, a walk and a goal towards which to go. If there is no goal we could not know what it is walking forward and what it is to going backwards. Therefore, to know whether science and technology progress it is necessary to agree on what are their goals, to measure whether we are closer or farther away from them.

So, if we advocate that the purpose of science is to discover the intimate structure of reality, to show us the world as it is, there would be many (instrumentalists) who would not say that science progresses, because in a few centuries we have gone from worldviews, for example in physics, which are radically incompatible (Ptolemy, Newton, Einstein) and that we cannot assure that they will endure. If on the contrary we propose that the objective of science is to explain and predict, understanding scientific theories as mere instruments of calculation, then we certainly can say that science advances as the number of facts explained increases, the predictions are more accurate and the precision of calculations is higher. In the instrumentalist point of view, science clearly progresses.

With respect to technology we can proceed in the same way. If the objective of technology is to provide human beings with control over nature, it is obvious that today we can control aspects of reality that were not available to us a few years ago. Therefore, in this sense, technology also progresses. One can speak without blush of scientific and technological progress.

What is perhaps more problematic is to presume that scientific and technological progress is equivalent to individual and social progress. In order to talk about the later kind of progress we should previously establish which are the objectives of individuals and the society. So, which is the goal of the human being? What is the meaning of human life? Obviously, the answer to these questions is far from unanimity. For many, the answer to this question could be that individuals seek happiness, although great schools of thought would not agree with this statement. Already in literature appear examples of how above happiness may be knowledge (recall the story of original sin) or power (the myth of Goethe's Faust). But even assuming that happiness is the goal of the human beings, are they nowadays happier than in older times? It seems at least doubtful to be able of doing such a strong claim.

If now we focus our attention on social progress, we find ourselves with the same difficulties and conditioning factors: what is the objective of a society? Is it to support the biological success of a species? In this case human societies, assisted by science and technology, have clearly progressed because the biological success of our species is undoubted. But perhaps other may demand societies other tasks such as equity among its members, peaceful and fair resolution of conflicts, protection of the weak, etc. And in this sense we cannot be sure that our modern western societies are more fair, equitable and protective than the primitive societies of hunter-gatherers were. Therefore, we see that, in both individual and social senses, the term progress is misleading and depends on the meanings assigned to individuals and societies. Scientific and technological progress are dissociated from the progress of individuals and societies (or at least not univocally related).

But, even understanding that it is totally debatable, we would dare to contribute to this debate with our understanding about the objectives of individuals and societies and thus be able to determine if there is progress in these areas or not. We propose that the objective of human life, of each individual life, the goal of any human, is his own humanization. Individuals are born to biological life, but they are made, they are built as human beings, partly by the influences of their environment and partly by their own efforts. Persons, to a certain extent, build themselves, they self- determine, ask themselves and respond to the fundamental question "what I want to be". And this human self-making is performed integrating its biological aspects, but overcoming them at a level of greater autonomy, that is, a level in which individuals increasingly give themselves their own norms, depend less on nature and society, they are more their own master.

Continuing this argument the objective of society would be to help the individuals to achieve their humanization, their autonomy. Then, agreeing that the previously mentioned objectives are valid, it can be asserted that indeed there is an individual and social progress. More and more individuals nowadays, and humankind as a whole, are less subject to nature and their own biological roots, having achieved that liberation largely by the help of science and technology. Likewise, current societies, which have fostered the development of science and technology, also increasingly enable individuals to give themselves their own rules. Therefore, even with the great limitations that still remain, we can still believe today in the progress of man and society, and in the fact that these have been largely fostered by advances in science and technology.

To illustrate this fact let us run a *gedankenexperiment*, a thought experiment. Let us imagine that most of us, university scholars of the highly developed world, would probably be at the beginning of last century (about 100 years ago) people with a life expectancy of 47 years[1] (33 for US black men) and without female vote (granted in 1920 in USA); 1,000 years ago ignorant farmers in the lands of our feudal lord (those with European ancestor); 10,000 years ago we would not know agriculture and we would roam in small bands to hunt and gather wild fruits; 100,000 years ago we would not have language yet with the corresponding limitations in the ability to think; a million years ago we would be a *homo habilis* individual of 1 meter tall and 600 cc of cranial capacity (1,400 cc current) that we would not have finished the biological evolution towards the contemporary human and 10 million years ago we would not differentiate ourselves from the gorillas and chimpanzees of that time.

## B. The faith in an endless progress

The conceptual paradigm in scientific research —and probably in the whole spirit of our times— it is assuming some kind of endless *progress*. But is this hypothesis supported by the facts? What is real and what myth in our faith in a continuous progress? Let us address this issue beginning with a tale.

A renowned publishing house celebrated the 50th anniversary of its foundation and, to commemorate it, decided to make a special edition of the Constitution of their country. For such a great event, the best specialists, notorious illustrators and leading graphic designers were hired. They chose high quality paper and luxurious bindings. So, to avoid any failure, the editor put all her effort to guarantee that there was not a single misprint throughout the book. She was aware of the difficulty of the mission, since she had many years of experience in the editorial business, and knew how, at the slightest opportunity, a typo in the text slipped. Therefore, she surrounded herself with the best team of editorial editors and hired a second external team among the best on the market. Not happy with that, the final result was submitted to a panel of independent experts and, finally, the editor in person reviewed each single word on each page, until she was sure there was absolutely no error. At last the copys were printed and their commercialization began. So proud was the company of the result of the work, that wanted to make clear to the possible clients the editorial effort that have been made so, to each copy, added a flap with the following inscription: "This edition does not contain any tipo" (yes, " tipo " instead of typo).

---

[1] *Source:* U.S. Dept. of Commerce, Bureau of the Census, *Historical Statistics of the United States.*

The present contribution, as it is obvious, does not deal with the problems of editing books, although the preceding story can introduce us to the question. Many people, and certainly many engineers can undoubtedly identify themselves with this anecdote. Man times, we make every effort to perform a task in the best possible way, taking care of all the details, and yet, any unforeseen failure override all or part of the work.

This situation, which in daily life can suppose a greater or lesser level of personal frustration, acquires another importance and different significance when it arise in a technological task. If, in the professional practice of engineers, the design or manufacture of a certain device fails, the consequences can be very serious or even irreversible. Modern societies have so highly trusted and depend on technology, that its failure can cause enormous material and economic damage, not to mention the possible loss of human lives. Therefore, the quality of technological products is a social requirement that, fortunately, is increasingly introduced into the culture of practicing engineers. They have to look for new more reliable technologies and check the quality of existing ones. It is true that even today, unfortunately, aviation accidents still occur, but there is no doubt that much less frequently than a century ago. It could be possible continue improving flight safety and get an aviation completely accident free? We will try to answer this question in what follows.

Perhaps an initial optimism, based on the confidence that society places in technology, will lead many to answer affirmatively. The answer, however, is more complex. Western thought has witnessed how solid conceptual buildings have surrendered to the erosion of time and a finer analysis. In Antiquity and in the Middle Age scientific knowledge in the West was expressed, among other doctrines, mainly through Aristotelian Physics and Philosophy. At the beginning of the Modern Age, however, this consolidated concept of the world suffered two devastating attacks. On the one hand, Galileo, among others, introduced experimentation as the most decisive criterion of corroboration, and the demand for a mathematical expression of the contents of knowledge: Modern Science emerged. On the other hand, Descartes, with his methodical doubt, inaugurated a new way of doing philosophy with respect to the preceding stages, which has shaped the modern mentality and which in many aspects rejects the previous methodology and philosophical thinking.

There is a lot of security that has been lost, but at least Science remained as a rock to grasp and modern reason as the maximum criterion to reach the truth. In this era great progress was made and it fills humanity with pride and optimism, perhaps living its peak moment in the period of the Enlightenment (18th century). Unfortunately, this situation ended in the late nineteenth and early twentieth centuries, when scientific discoveries are made that, on the one hand, seriously question the image that science had of the world until that moment and, secondly, put limits to scientific knowledge itself. The theory of relativity appeared, which revolutionized primary concepts such as those of space and time, and quantum mechanics that limits what we can know about a particle (Heisenberg's uncertainty principle). And even in apparently sure fields such as mathematics, when reviewing the founding principles (Hilbert and Russell among others) a loss of certainty arose[2] and the mathematical knowledge is irreversibly self-limited (Gödel's theorem).

Problematic scenery for the man of the 21th century: neither Philosophy nor Science can give us security, then, where will we look in this search for certainty? Fortunately Technology still remains. Science may not explain the world in a sure way, but at least its predictions work, and with them you can build all kinds of artifacts that improve our place in the world. The society of our century relies its trusts in Technology. It is true that technological advances were questioned as the result of the terrible results of the First and Second World War, but today that is an old-fashioned debate. Technology works, and it works very well. We keep it in mind at every moment of our daily existence and, despite the criticism of recent years, that man has already been installed in a technological world. Moreover, nowadays certain applications and ways of using technology may be questioned, but nobody seriously proposes a turnaround, a return to a pre-technological situations.

## C. Limits of technology

Let's start this section with a few press references.
- "U.S. Customs officials said Tuesday that they had traced the source of last weekend's system outage that left 17,000 international passengers stranded in airplanes to a malfunctioning network interface card on a single desktop computer in the Tom Bradley International Terminal at LAX[3]".
- "In what appears to be the first pedestrian fatality involving a self-driving car, an Uber vehicle operating in autonomous mode Sunday night struck a Tempe, Ariz., woman, who later died of her injuries at a local hospital[4]".
- "Software failure caused $1.7 trillion in financial losses in 2017. Software testing company Tricentis found that retail and consumer technology were the areas most affected, while software failures in public service and healthcare were down from the previous year[5]".

We all have experience that, even when we put our best effort to perform a task in the best possible way, taking care of all the details, however, some unforeseen failure spoils all or part of the work. This situation, which in daily life can suppose a greater or lesser degree of personal frustration, acquires another importance and different significance when it comes to a technological task. If, in the professional practice as engineers, the design or manufacture of a certain device fails, the consequences can be very serious or even irreversible. Contemporary societies have relied on and depend to such a high degree on technology, that its failure can cause enormous material and economic damage, not to mention the possible loss of human lives. Therefore, the quality of technological products is a social requirement that, fortunately, is increasingly introduced into the culture of practicing engineers. They have to look for new more reliable technologies and check the quality of existing ones. It is true that even today, unfortunately, there is an aviation accident, but there is no doubt that much less frequently than at the beginning of the century. Can you continue to improve and get aviation completely accident free? We will try to answer this question in what follows.

---

[2] Ernest, P. (2016). The problem of certainty in mathematics. *Educational Studies in Mathematics*, *92*(3), 379-393.
[3] Los Angeles Times, August 15,2007
[4] San Francisco Chronicle, March 19, 2018
[5] TechRepublic, January 26, 2018

In the field of building computer systems, a subject that we have studied with special interest and in which we have experience from our professional practice, we are aware that any system contains an undetermined number of undetected errors. Precisely for this reason, we try to advance by getting more and more reliable products by improving the technique of system construction, having created a discipline that, under the name of "Software Engineering", brings together a set of techniques aimed at producing software quality. Several authors have pointed out that the number of errors detected by a quality control process is greater when we increase the effort devoted to this surveillance task. This relationship works well at the beginning but, nevertheless, at some point an increase in the effort dedicated to quality control does not translate into a greater detection of errors, leaving always a remaining number of undetected errors.

For other authors, the process is even more devastating, since the improvements that are achieved through quality control are compensated by the new errors that appear when introducing modifications in the original software. Therefore, the evolution over time of the number of errors presents a first downward behavior due to quality control, followed by a subsequent increase due to changes in the software.

This low quality of software developments means that the average duration of computer projects grows alarmingly as their size increases, as reflected in the graph. Even in many of these projects, the failure can be so spectacular that they even may result canceled, with the loss of all the investment made. The probability of failure and, therefore, of cancellation, also increases with the size of the system.

But the size and complexity of the computer system does not stop growing. It is estimated that the amount of code installed in most consumer products (televisions, washing machines, etc.) is doubling every two years. New and better programming techniques struggling against a greater software complexity. The question arises immediately: will it be possible to continue improving and to obtain software free of errors?

Without questioning the validity of Technology and its central role in contemporary society[6], we do want to ponder its limitations. Faced with unquestionable successes and progress, daily failures occur. Can these failures of Technology be overcome? Is the Technology solution more and better Technology? In our opinion, clearly not. There are fundamental reasons that limit the technological capacity of humans. Frequently, faced to the failures, it is normal to look for an external cause: fatigue of materials, extreme environmental conditions, human negligence, etc. Without denying these realities, the first source of many of these failures must be sought in more depth. In this sense, failures in the software can be paradigmatic, since they can hardly be attributed to material causes. In our opinion, the deep origin of many failures should be placed in an intrinsic limitation of technological knowledge.

Technology is limited, at least, by three factors: a) Material limits; b) Scientific limits; and c) Operational limits. Let us study this topic in a more detailed way.

## a) Material limits

The first and most obvious limitations of technology are those imposed by materials. You cannot get materials more resistant, or lighter, than those available at a given historical moment. And whatever these materials are they will have limits of operation due to temperatures, pressures, corrosion, fatigue, etc. For example, one of the fundamental

---

[6] Ramón Queraltó: *Mundo, Tecnología y Razón en el fin de la Modernidad*, Barcelona, P.P.U., 1993

limitations that prevent the development of fusion energy is the unavailability of suitable materials to build the reactor.

Within the field of information technology we can also point out some material limits. For example, the construction of more powerful computers involves increasing the complexity of the circuits that compose them, that is, by increasing the number of transistors per chip. To do this, we must continue advancing in miniaturization, making each individual transistor measure less and less. But we find a first limitation, which is not too far, due to the wavelength of De-Broglie. Below these dimensions the electrons stop behaving like classical particles and quantum phenomena must be taken into account. A second limitation arise when we consider that it is difficult to wonder how , with current technology, a transistor could be made to measure less than an atom.

### b)   Scientific limits

The second limitation of technology is given by the limitations of scientific knowledge that should support it. These limitations may be due to a specific historical moment or intrinsic to scientific knowledge. Thus, for example, many diseases have been cured when it is known scientifically what his cause. For example, many stomach ulcers are cured today with antibiotics to identify that one of its causes is a certain bacteria. On the other hand, meteorological technologies (prediction and modification of climates) has to deal with the chaotic nature of a large part of the models of air physics and the limits that this fact impose on the technologies.

### c)   Operational knowledge limits

But sometimes technology fails and it is not due to any of the two previous causes. For example, in Computer Science there is a saying that states that in life there are only three certainties: death, taxes and errors in the programs. These errors cannot be attributed to material defects, because there is no material in software, or to scientific problems, because the bases of the discipline are well established. The conceptual complexity of computer systems means that their implementation, the operational hands-on process, introduces errors. This is the intrinsic limitation of the technological knowledge to which we referred earlier and on which we are going to focus.

What are the causes of this limitation? It is not possible to think that it is due to an ignorance of the basic principles that operate in the phenomena that rule the artifacts. We can suppose that the scientific knowledge that gives support to the design and construction of an artifact is correct and nevertheless failure of the artifact can occur. Technological knowledge is basically operational, it is not so much a *know-what* as a *know-how*. And is this know-how, in our opinion, necessarily limited, which causes errors and failures. You can know perfectly, for example, the logical principles that support computer programming, and apply the most sophisticated software engineering techniques, but nevertheless errors will still exist. And as the size and cognitive complexity of the artifact increases, the situation gets worse.

Now, why think that the situation will not be corrected when a better Technology is obtained? What reasons support our assertion of the limitation of technological knowledge? We can identify at least three arguments, each of them in a different plane and, consequently, with different argumentative strength.

The first of them, that we call the historical argument, is supported by a double reasoning. In the first place, the historical argument gets support from a certain logical coherence. In fact, if previously Philosophy and Science had promised us a security that they could not provide us, why should we trust now in Technology? In this sense, the burden of proof would fall on those who state that Technology will stop failing with more Technology. Second, the history of Technology is full of failures that reach our days, which indicates more in the way of the limitation that we propose than in the opposite direction. This argument can be criticized as being a circumstantial evidence, although it does not lose its demonstrative force as a very probable historical presumption. However, the other two arguments are perhaps more radical.

Let's start with what we call *the recursion argument*. A human being has been defined as a *being-in-the-world* (Heidegger) or a *being formally installed in reality* ( Zubiri ), and this reality is already given to human, imposed on him/her, overflows him/her, and that human is part of reality. So, an unlimited *know-how* about reality must include human himself/herself, which is an absurdity. The capacity for technological action must be limited so as not to be contradictory. Otherwise, we would find ourselves in the same circumstance that Escher masterfully traces in his lithography *Drawing Hands*, where he depicts a right hand which is drawing a left hand that, in turn, is drawing the right hand that draws it. Even if someone has a perfect representational knowledge of the world (*knowing what*), no strategy (*know how*) can be thought that substantively alters (act upon) the subject of operational knowledge and, additionally, describes the process and results of the action. For example, an isolated subject cannot design a strategy that allows him/ her to describe the sensations of a completed suicide. Analogously, an isolated robot cannot design a strategy to describe the state of its circuits when it is turned off, although it can design them to turn off and reconnect.

This argument, in effect, imposes a limit on technological knowledge, but it could be argued that, except in the cases described, it is of little practical interest, and that, consequently, the technological society has no practical limits in its *know-how*. Let us see, however, that this is not the case through the third and last of the arguments, which is what we call *the complexity argument*. The technological subject (be it an individual or a society) has a *know-how* that projects in a technological action. This process therefore implies knowledge of a technological nature, which resides first in the subject and then is reflected in reality. Therefore, the complexity of the technological action carried out cannot be superior to that of the *know-how* of the subject. We understand here by complexity the amount of information necessary to describe a technological action or *know-how*. But, on the other hand, the complexity of *know-how* cannot surpass that of the technological subject. If, according to Shanon's information theory, we measure the amount of information in bits, we can formulate the complexity argument by saying that the number of bits describing a technological action cannot be greater than the storage capacity and process of the technological society that carries it out. And, as the storage capacity and process of any technological society, present or future, will necessarily be limited, so will its *know-how*, its technological knowledge. As the complexity of a technological action is approaching to the capacity of the subject of that action, the number of aspects that are not taken into account grows. As a consequence of this, failures will be more frequent.

## D. The inoperancy principle

In light of the approaches described, and based on these three arguments, we propose the *inoperancy principle* of Technology resembling the uncertainty principle of Physics. This principle can be formulated in the following way: the complexity of a technological action and its mean time between failures (MTBF) are inversely proportional, where the constant of proportionality depends on the technological capacity of the subject that performs the action. Obviously in this expression, when we talk about failures, we are referring exclusively to those due to a lack of *know-how*, not to defects in materials, wear, fatigue, etc. On the other hand, in this formulation, the term of inverse proportionality does not refer to a certain mathematical function, but rather describes a qualitative rather than a quantitative behavior.

## Acknowledgement

I thank Hezheng Yin (UC Berkeley) for assistance with the text editing and comments that greatly improved the manuscript.

## Funding sources

This work has been financed under the Framework Agreement between Andalucía TECH, VLC CAMPUS and the University of California (Berkeley). Activity financed within the framework of the Campus of International Excellence program of the Ministry of Education, Culture and Sports, Spain.